# Assessing sources of uncertainty in length-based estimates of body growth in populations of fishes and macroinvertebrates with bootstrapped ELEFAN

R. Schwamborn, T. K. Mildenberger, M. H. Taylor

Oceanography Dept., Federal University of Pernambuco (UFPE), 50670-901 Recife, Brazil.
e-mail: rs@ufpe.br

National Institute of Aquatic Resources, Technical University of Denmark, Kemitorvet, 2800 Kgs. Lyngby, Denmark

Thünen Institute of Sea Fisheries, Herwigstraße 31, 27572 Bremerhaven, Germany

## Abstract

The determination of rates of body growth is the first step in many aquatic population studies and fisheries stock assessments. ELEFAN (Electronic LEngth Frequency ANalysis) is a widely used method to fit a growth curve to length-frequency distribution (LFD) data. However, up to now, it was not possible to assess its accuracy or the uncertainty inherent of this method, or to obtain confidence intervals for growth parameters within an unconstrained search space. In this study, experiments were conducted to assess the precision and accuracy of bootstrapped and single-fit ELEFAN-based curve fitting methods, using synthetic LFDs with known input parameters and a real data set of *Abra alba* shell lengths. The comparison of several types of bootstrap experiments and their outputs (95% confidence intervals and confidence contour plots) provided a first glimpse into the accuracy of modern ELEFAN-based fit methods. The main components of uncertainty (precision and reproducibility of fit algorithms, seed effects, sample size and matrix information content) could be assessed from partial bootstraps. Uncertainty was mainly determined by LFD matrix size (months x size bins), total number of non-zero bins and the sampling of large-sized individuals. A new pseudo-Rsquared index for the goodness-of-fit of VBGF models to LFD data is proposed. For a large, perfect synthetic data set, pseudo-Rsquared$_{Phi'}$ was very high (88 to 100%), indicating an excellent fit of the VBGF model. The small *Abra alba* data set showed a low pseudo-Rsquared$_{Phi'}$, from to 54% to 68%, indicating the need for more samples (length measurements) and a larger LFD data matrix. New, robust, bootstrap-based methods for curve fitting are presented and discussed. This study demonstrates a promising new path for length-based analyses of growth and mortality in natural populations, which are the basis for a new suite of methods that are included in the new fishboot package.

**Key words:** Body Growth; von Bertalanffy Growth Function; Bootstrapping; ELEFAN; data-poor fish stock assessment; fishboot package

**Highlights:**

- ELEFAN-based fit methods for the analysis of length-frequency data were tested and improved.
- The new, bootstrapped approach provides best fits and 95% confidence intervals for all parameters.
- This new approach showed a high level of reproducibility and accuracy.
- A new statistic for the information content of length-frequency data is introduced (pseudo-$R^2$).
- This study demonstrates a promising new path for length-based studies of aquatic populations.

1. **Introduction**

The estimation of body growth is crucial for the understanding of populations and ecosystems, and is the basis for all analytical stock assessment methods in fisheries science. Modal progression analysis, a length-based method, has been used to infer body growth of fish and aquatic macroinvertebrates since the very beginning of fisheries science (Petersen, 1891). In this approach, month-by-month time series of histograms of fish length (the monthly length-frequency-distributions, LFDs) are plotted and the peaks (modes) are connected to track the growth of each cohort from month to month (Fig 1). Usually, the 'von Bertalanffy Growth Function' (VBGF, von Bertalanffy, 1934, 1938) is fitted to such monthly LFD data. This simple approach, the so-called 'Petersen method', became immensely popular since the development of computer-based methods such as ELEFAN I (Pauly and David, 1981; Pauly, 1986; Pauly, 1987), that were later incorporated into the software packages 'COMPLEAT ELEFAN' (Gayanilo et al., 1987) and 'FISAT II' (Gayanilo et al., 1995, 2005). These methods were recently implemented in R, within two new packages called 'ELEFAN in R' (Pauly and Greenberg, 2013) and 'TropFishR' (Mildenberger et al., 2017a; 2017b), which also contain numerous other functions.

ELEFAN I (Pauly and David, 1981) uses a high-pass filter to identify peaks in LFDs. This filter is based on a moving average, where the frequencies of the original LFD that reach above the moving average are detected as peaks (black bars in Fig. 1) and those that are below the moving average are detected as troughs (white bars in Fig. 1). The score of peaks (black bars) that are being crossed by a VBGF curve (Fig. 1) is the basis for calculating a goodness-of-fit indicator ('Rn' score). Any

ELEFAN-based fit procedure is thus nothing else but the search for the single optimum combination of VBGF parameters that produces the best results i.e., the highest possible Rn value.

Although there are more accurate length-at-age or tagging-based methods available, this length-based approach is still extremely relevant, especially when resources and data are limited (e.g., under data-poor situations), and for organisms where tagging or ageing is simply not possible (e.g., for many small-sized shrimp species).

One fundamental problem during this fit procedure is that the VBGF parameters $L_\infty$ (asymptotic length) and K (growth constant) are interrelated, i.e., using higher values of $L_\infty$ always leads to lower estimates of K and vice-versa. The most widely used approach to circumvent this problem during the fit of the VBGF curve has been to fix $L_\infty$ *a priori* and then concentrate all effort in precisely determining K, e.g., by using a K-scan routine (Gayanilo el al., 1995). The two methods commonly used to fix $L_\infty$ from LFDs are the $L_{max}$ approach (simply using the length of the largest fish to assess $L_\infty$, e.g. Mathews and Samuel, 1990; Sparre and Venema, 1998) and the Powell-Wetherall plot (P-W plot, Wetherall, 1986; Wetherall et al., 1987). Recent in-depth simulations showed that both methods (Lmax approach and P-W plot) have fundamental issues and will produce severe bias in their $L_\infty$ estimates in many situations and should thus not be incautiously used and recommended (Schwamborn, 2018). Thus, it seems likely that it is not possible to fix or constrain $L_\infty$ *a priori*, just by looking at the largest fish or by analysing a catch curve with a P-W plot (Schwamborn, 2018). *A priori* fixing of $L_\infty$ has been suggested to be imperative when fitting VBGF curves to LFDs by many authors (e.g., Gayanilo et al., 1995; Sparre and Venema, 1998,), even when using recent length-based methods (Taylor and Mildenberger, 2017).

When $L_\infty$ is not fixed or constrained *a priori*, fit methods can be used that perform an unconstrained simultaneous search for an optimal combination of all growth parameters (at least for $L_\infty$ and K), within all possible combinations. A widely used method for such an unconstrained search is ELEFAN I with Response Surface Analysis (RSA, Gayanilo et al., 1987; Brey et al., 1988, Fig. 2). It consists in calculating the ELEFAN I goodness-of-fit indicator Rn for different combinations of K and $L_\infty$ within a pre-defined grid of discrete input parameter combinations. A heatmap (the RSA plot) is then plotted with $L_\infty$ and K as x and y axes and the Rn value as the heat colour (Fig. 2). The highest peak (i.e., the highest Rn value), is then presented as the optimum fit for a given LFD data set.

Two new ELEFAN-based curve fitting algorithms have been developed recently, that also allow for an unconstrained search within the multi-dimensional parameter space: ELEFAN_SA and ELEFAN_GA, which are part of the R package TropFishR (Mildenberger et al., 2017). ELEFAN_SA uses simulated annealing ('S.A.'), a common probabilistic technique for approximating a global optimum, that is included in the GenSA package (Xiang et al. 2013). ELEFAN_GA (Mildenberger et al., 2017) is a different, more complex and sophisticated VBGF curve fit application, which is based on a genetic algorithm ('G.A.'), a metaheuristic inspired by the process of natural selection (also known as evolutionary algorithm). It uses the GA package in R (Scrucca, 2013) to maximise the ELEFAN I goodness-of-fit score Rn. Both algorithms search for the best fit through a series of iterations (consecutive curve fitting attempts) and allow for unconstrained search for optimum combinations of growth parameters, including seasonality.

Similar to RSA, these methods produce only a single output, i.e., one optimum combination of "best" parameter estimates, without providing 95% confidence intervals (CIs) for any estimates. Furthermore, it is possible that these two methods may be prone to the attraction effects of local maxima, where the strength of these effects will depend on algorithm settings used. However, this potential pitfall has not yet been analysed in depth in any study, and no measures have been yet proposed that could systematically circumvent problems due to the attraction to local maxima.

Accuracy of length-based methods and their susceptibility to local maxima has been discussed since many decades (Morgan and Pauly, 1987; Kirkwood and Hoggarth, 2006; Pauly and Greenberg, 2013; Taylor and Mildenberger, 2017). The occurrence of local maxima is a particularly dangerous phenomenon, since their effect can hardly be estimated in routine analyses.

Despite the widespread use of length-based methods, there have not yet been made any efforts to quantify the uncertainty in growth parameter estimates derived from to sample size (total N), LFD matrix size (months x size bins), LFD data structure (too few samples, biased or irregular sampling, non-VBGF growth curve shapes, no clear cohorts, few or no large individuals, etc.) or due to the behaviour of the fitting algorithms (e.g., non-convergent behaviour, seed effects, attraction to local maxima).

The starting point for the analyses presented in this study was thus to analyse the susceptibility of the new curve fitting algorithms to be captured by local maxima, by applying them repeatedly to the same data, but using different starting points for the search (i.e., different 'seed values'). Such an experiment of multiple repeated fit attempts may be considered as a partial bootstrap, since it will reveal only a part of the total uncertainty, i.e., the uncertainty due to the choice of starting points for randomized algorithm search processes (i.e. "seed effects"). Any variability shown in such experiments will only show the errors induced by the fit procedure and seeds effects can be visualized in plots and quantified, but not the uncertainty due to stochastic errors during acquisition of length samples in the field.

In theory, an ideal fit algorithm should always be able to find the overall best growth model, irrespectively of the initial values used for the search. However, in practice, even the best fit algorithm may be susceptible to get stuck in a local maximum, and it may thus be far from trivial to find or even to define the exact location of such an overall maximum within a complex multi-dimensional search space. Instead, a more appropriate solution may be to provide a range of likely best fits, or more precisely, the range where the underlying mean parameters of the population most likely are located (i.e., the CIs of the parameter estimates). For complex statistical problems, where the underlying distributions cannot be known *a priori*, this is usually obtained by full nonparametric bootstrapping (Efron, 1979; Efron and Tibshirani, 1993), which has become the standard approach for the estimation of the uncertainty or error of numerous methods and models, since the emergence of fast computers (e.g., Halfon, 1989; Mocq et al., 2013; Radinger et al., 2017). Bootstrapping is especially useful where accurate analytical expressions for error terms cannot be easily obtained (e.g., Sohn and Menke, 2002), or where there is a complex non-linear behavior of multiple interacting parameters, which is the case in any VBFG curve fit procedure.

The next step was thus to develop a new bootstrap method (see details below) to assess the overall uncertainty inherent of the new fit algorithms ELEFAN_GA and ELEFAN_SA and to calculate proper CIs based on the empirical quantiles of the bootstrap posteriors (Efron, 1979; Efron and Tibshirani, 1993).

The original paper describing ELEFAN I by Pauly and David (1981) stated that " … searching for the optimal combination of four parameters … can become elephantine". Here, we present an endeavour to search for an optimum VBGF curve fit through multiple repeated fit attempts, and even more, a full bootstrap method that provides nonparametric CIs for each VBGF parameter.

This study intends to show that this "elephantine" challenge can be met in a robust and reproducible way and goes beyond this, presenting a bootstrap-based method, i.e., a practical tool for general use by scientists and fisheries managers.

## 2. Materials and Methods

### 2.1 Data sets

A natural LFD data set (*Abra alba*, Brey et al., 1988) was used to test the reproducibility (e.g., seed effects and local maxima) and precision of length-based methods when applied to real LFD data with a small sample size (Fig. 1). The 'alba' data set is composed of shell lengths of the white clam *A. alba* (W. Wood, 1802). These are the shell lengths of 675 *A. alba* individuals collected over seven sampling months and binned into 14 length classes. This data set was chosen among the data available in TropFishR because of its small size, which allowed very fast repeated analyses. Also, its simple, noticeably deficient structure (Fig. 1) was used to contrast with the perfect synthetic data set.

A synthetic LFD data set ('synlfq5b', 12 months, 36,711 individuals, K = 0.5, $L_\infty$ = 80) was used to test the accuracy and reproducibility of ELEFAN-based fit methods (Fig. 1). The synlfq5b dataset was created by selecting the first 12 months from the original synlfq5 data set in TropFishR, in order to increase the speed of analysis.

*2.2 Curve fitting methods tested*

Three curve fit methods were analysed in this study, all of which are part of the TropFishR package (Mildenberger et al., 2017): ELEFAN (includes RSA, not being used to produce CIs), ELEFAN_SA (used for bootstrapping) and ELEFAN_GA (used for bootstrapping). All of them use the same ELEFAN I - based principle of curve fit (Pauly and David, 1981; Taylor and Mildenberger, 2017). During each fit procedure, a goodness-of-fit estimator ('Rn') is calculated, based on the number or positive peaks that are crossed by the VBGF curve. In TropFishR, all three fit algorithms (ELEFAN with RSA, ELEFAN_SA and ELEFAN_GA) use the same 'lfqFitCurves' function to calculate Rn, but they differ in the method used to adjust the best growth model, i.e., in the algorithms used to find the single combination of VBGF parameters that gives the maximum Rn value.

ELEFAN with RSA uses the simple, well-established Response Surface Analysis plot, where the goodness of fit Rn value is displayed in a heatmap plot composed of N discrete cells within a grid of K *vs* $L_\infty$. After calculation of 'N' Rn values and plotting the heatmap, the cell with the highest

Rn value is then presented as the best parameter combination in the basic ELEFAN function in TropFishR. ELEFAN_SA (Mildenberger et al., 2017) uses common simulated annealing (Xiang et al., 2013), while ELEFAN_GA (Mildenberger et al., 2017) is a more complex VBGF curve fit based on a genetic algorithm (Scrucca, 2013).

Speed and precision can be finely tuned within these fit algorithms, which have several input settings. For RSA, the key parameter that defines speed and precision is the resolution of the plot (i.e., the size of the input grid of K *vs* $L_\infty$). For the ELEFAN_SA and ELEFAN_GA fit procedures, all settings were optimized for maximum precision. Settings for ELEFAN_SA were chosen to ensure the widest possible search of for the overall maximum (high temperature value) and a very large number of iterations, to ensure full convergence of iterations towards one best fit value (i,e., high maxit value, to avoid the occurrence of any incomplete fit runs), regardless of the necessary computation time (maxit = 500, SA_temp = 5e+05, SA_time = 4 minutes, Schwamborn et al., 2018b).

For the more complex genetic algorithm ELEFAN_GA, three key settings were fine-tuned for precision (Table 1). Settings for precision-optimized ELEFAN_GA were maxiter = 50, run = 10, pmutation = 0.2, for both datasets, with popSize = 100 for the *A. alba* satset and popSize = 60, for the synlfq5b dataset, and default TropFishR settings for all other parameters. The search space used during fit optimization was always constant with $t_0$ = 0 to 1, $L_\infty$ = 8 to 15, K = 0.1 to 5, for *A. alba*, and $L_\infty$ = 60 to 120, K = 0.05 to 2, for synlfq5b. Moving average span (MA) for curve fits was always set to MA = 9 for the *A. alba* dataset, and MA = 11 for the synlfq5b dataset (Taylor and Mildenberger, 2017).

*2.3. Design of bootstrap experiments*

First, a series of repeated VBGF curve fits was conducted, which may be called 'partial bootstrap' simulations (PBoot). In PBoot experiments, all runs were performed with the same, original data, but with different random seed values. Thus, in PBoot experiments, any differences between output parameter estimates are only due to the imprecision and lack of reproducibility of the optimization-based fit method used (i.e., vulnerability of the fit algorithm to seed effects and local maxima). The PBoot routine assessed only the precision of the fit method, under the initial assumption that the data perfectly represent the actual population (zero error due to sampling in the field). It does not consider the N of sampled animals. It just measures the reproducibility of the results for a given fit algorithm and a given LFD data set.

Then, a full bootstrap (FBoot) was conducted, with random resampling from LFDs, as to assess the uncertainty in sampling and fit. The FBoot routine considers sample size and sample structure. It considers variability in the population and uncertainty in sampling the animals, considering the N of sampled animals and the accuracy of the method. The main difference between PBoot and FBoot is that FBoot considers all this, while PBoot measures only the precision of the fit method.

The relatively long time (several minutes) needed to build a single high-resolution RSA plot precludes its use for standard bootstrap routines. Thus, in this study, bootstrap analyses were conducted with the fit algorithms ELEFAN_SA and ELEFAN_GA. These two fit algorithms were applied to both data sets (*Abra alba* and synlfq5b).

The FBoot procedure consists of the following three steps, based on a set of monthly LFD histograms (for 1,000 iterations, repeat steps A and B 1,000 times):

- A.) Resample with replacement from each monthly LFD sample (Fig 3) and build a new, complete LFD data set with these random samples.
- B.) Fit a VBGF curve to these LFD data (using ELEFAN_GA or ELEFAN_SA) and record the parameter estimates.
- C.) Determine 95% quantiles for all parameters (i.e., bootstrap 95% CIs).

Generally, 1,000 runs were conducted for the bootstrap experiments (PBoot and FBoot). For each run within each loop, a unique new global seed value ensured full stochasticity of simulations within the R environment.

CIs of the VBGF output parameters ($L_\infty$, K, $t_0$) and for the growth performance index Phi' (Phi' = $\log_{10}(K) + 2 * \log_{10}(L_\infty)$, Pauly, 1979; Pauly and Munro, 1984) were calculated by using the 95% quantiles of the posterior distributions. The relative magnitude of uncertainty due to seed effects (i.e., fit method artefacts) and due to the data structure were evaluated by comparing the CIs in partial (PBoot) and full bootstraps (FBoot), for the two fit methods (ELEFAN_SA and ELEFAN_GA) and for the two example data sets (*Abra alba* and synlfq5b). Thus, a total of 2 x 2 x 2 = 8 bootstrap experiments were conducted, with 1,000 runs each, totalling more than 8,000 fit optimization runs (Exp 1.1 to 4.2, Tables 2 and 3). Furthermore, high-resolution (100 x 100) response surface analyses (RSA) of these two data sets were conducted (using the standard ELEFAN function in TropFishR), with varying grid settings, for comparison.

*2.4 Comparing posterior distributions*

To test for significant differences between quantiles and medians of the posterior distributions (i.e., differences between two distributions with regard to the position of the lower and upper 95% quantiles) obtained with different fit methods and bootstrap routines, a non-parametric Harrell–Davis quantile test was conducted at alpha = 0.05, using the function 'Qanova' within the R package WRS2 (Mair and Wilcox, 2017).

The quantile test can be applied to compare two samples regarding the location of each specific quantile (e.g, quantiles of 0.025, 0.975, median), but it does not allow for a direct comparison of interquantile ranges. Such comparisons of inter-quantile ranges widths between PBoot and FBoot routines and between fit algorithms were done by standardizing (i.e., transforming) the posterior distribution of estimates prior to quantile tests. Standardization was necessary to eliminate the effect of location (e.g., differences in medians), when comparing inter-quantile ranges. Standardized estimates were calculated as "Standardized estimates = raw estimates – lower 95% quantile", for each element in each posterior distribution. Quantile tests were then used to compare the upper 95% quantiles of standardized $L_\infty$ and K values, to test for differences in inter-quantile ranges, using the R function Qanova at alpha = 0.05.

*2.5 Testing for reproducibility of the bootstrapped fit methods*

Reproducibility of the new, bootstrap-based fit methods was tested by repeatedly performing the whole bootstrap analyses with the same data, using the same fit parameters and the same fit algorithm (e.g., ELEFAN_GA). All bootstrap analyses were compared regarding their key outputs (posterior distributions), i.e., lower, upper 95% quantiles, inter-quantile ranges, and medians for K, $L_\infty$, and Phi', using quantile tests.

*2.6 Components of uncertainty*

The comparison of two types of bootstraps (PBoot and FBoot) and their outputs (CIs and confidence contour plots) allows an evaluation of modern ELEFAN-based fit methods and an assessment of the relative magnitude of errors from specific sources. The magnitudes of CIs calculated from PBoot experiments are determined by the precision of the fit algorithm (e.g., its susceptibility to seed effects). CIs obtained by the FBoot routine represent the total error, or total uncertainty, of the whole procedure, beginning from the stochastic error introduced when sampling in the field to the final curve fitting. These full CIs are determined by the same error that affects the PBoot CIs, but also by sample size (total and monthly N individuals), and by the growth-related information content (GRIC) of the LFD data matrix, in an additive way. GRIC can be composed by many factors, such as matrix size (number of sampling months, size bin resolution, length range), and LFD shape (number and distribution of non-empty size bins, presence of modes and of modal progression, number of individuals in the smallest cohort, etc.).

Thus, the ratio of confidence interval widths (CIW) from PBoot and FBoot experiments can be used to calculate a practical index of data quality for LFD data and their suitability for growth models, the Matrix-Information-Effect index, or pseudo-$R^2$ ($pR^2$):

$pR^2 = CIW_{PBoot} / CIW_{FBoot}$ ,

where $CIW = CI_{upper} - CI_{lower}$.

Pseudo-$R^2$ values can be calculated for each VBGF parameter and will generally range between zero and 1. Pseudo-$R^2_{Phi'}$ values close to 1 indicate highly informative LFD data with excellent VBGF curve fits. Pseudo-$R^2$ was calculated only for experiments with precision-optimized fit settings (i.e., maxiter = 50, etc.), to avoid potential bias due to incomplete fit (i.e., to avoid the occurrence of incomplete fit runs, where iterations still did not converge).

All calculations, experiments and statistical analyses were carried out in R (version 3.4.2, R Development Core Team, 2017) using its standard functions and the packages TropFishR (Mildenberger et al., 2017, version 1.2.1), WRS2 (quantile test, Mair and Wilcox, 2017, version 0.3-2), and ks (kernel smoothing for confidence contours shown in graphs, Duong, 2018, version 1.11.0). All R scripts required to reproduce the analyses are available in the Supplementary Materials (Schwamborn et al. 2018b, Supplementary Material: https://doi.org/10.6084/m9.figshare.5977840). The new bootstrap functions used in this study are available within the new 'fishboot' package (Schwamborn et al., 2018a).

## 3. Results

*3.1 Response surface analysis (RSA)*

High-resolution response surface analysis (RSA) always produced informative plots with extensive "basins" (vast areas with low Rn values) and discrete plateaus (Figs. 2 and 6). All RSA plots revealed the existence of numerous local maxima (secondary peaks), distributed along a characteristic "banana-shaped plateau" (Figs. 2 and 6). The overall shape of the plateau is generally consistent for a given data set, but the relative height and relative distribution of the numerous small peaks varies strongly with grid design, leading to different best-fit estimates for each arbitrarily chosen input grid of $L_\infty$ and K values.

For the synlfq5b data set, two attempts to find the best-fit VBGF curve with high-resolution RSA (100 x 100) grids produced considerably different best fit parameters, only because the input K and $L_\infty$ ranges of the grid were slightly modified. The best fit (highest Rn value) obtained from the first RSA, with input K varying from 0.1 to 0.8 $y^{-1}$ and input $L_\infty$ from 50 to 150 cm (100 x 100 grid) was $L_\infty = 105.0$ cm, $K = 0.28$ $y^{-1}$ (Fig 2). These estimates were very far away from the "true" initial parameters that had been used to generate these synthetic data ($L_\infty = 80.0$ cm, $K = 0.5$ $y^{-1}$). The original parameters were not even approximately close to the best fit and not within the "banana-shaped plateau" (Fig 2), indicating an overall low level of accuracy and reproducibility for this method, even when using a very high resolution.

Conversely, the best fit (highest Rn value) from another RSA, with a slightly different grid, with input K varying from 0.1 to 4 $y^{-1}$ and input $L_\infty$ from 40 to 160 cm (also in a 100 x 100 grid) was very precise, the best fit being virtually identical to the original "true" parameters: $L_\infty$ = 80.0 cm, $K = 0.49$ y$^{-1}$ (Fig 6).

Moreover, this method was very time-consuming, since VBGF models for all possible combinations of K and $L_\infty$ have to be analysed one-by-one, and there is no parallel processing with RSA. Most RSA plots took more than 20 minutes on a common laptop with a dual-core i5 processor. This slowness precluded its use for bootstrapping.

*3.2 Partial bootstrap (PBoot) experiments*

Both optimisation-based fit algorithms tested (ELEFAN_GA and ELEFAN_SA) showed to be extremely fast and practical, being several times faster than high-resolution RSA. Both fit algorithms were much faster than high-resolution RSA.

PBoot experiments showed that both fit algorithms (ELEFAN_GA and ELEFAN_SA), when used as one-time fit applications, display a very low level of replicability (single dots in Figs. 4a, 4c, 5a and 5c). Each repeated fit generally produced different results, even when using the same algorithm and the same search settings. For example, the range for $L_\infty$ estimates obtained for the sylfq5b dataset varied from 64.4 to 119.0 cm (ELEFAN_GA, maxiter = 50). This showed a considerable bias in $L_\infty$, ranging from -24% to +34%, using exactly the same search settings and the same data.

The variability of the outputs obtained when applying these algorithms repeatedly to the same data was quantified in the PBoot experiments (Fig. 4 to 6, Tables 3 and 4). The darker coloured areas of the scatterplots in Figs. 4 and 5 often reflect multiple overlying points, i.e., local maxima that were recurrently hit by these fit algorithms, using the same input data, but starting from different random initial seed values. ELEFAN_GA was much less prone to become stuck in exotic, extraneous local maxima than ELEFAN_SA. ELEFAN_GA became stuck in a discrete set of generally less than 30 local maxima.

Outputs obtained with ELEFAN_GA were generally significantly closer to each other than those obtained with ELEFAN_SA (Harrell–Davis quantile test, Tables 3 and 4). The 95% interquantile range for ELEFAN_SA was up to 141% larger than for ELEFAN_GA (Table 4), showing a much higher precision of the ELEFAN_GA method, which is based on a highly complex genetic algorithm. The only exception was the estimation of K for the synLFQ5b dataset, where CIs obtained with ELEFAN_SA were not significantly larger (PBoot) or even slightly (7%) smaller than CIs obtained with ELEFAN_GA (Table 4).

This set of experiments disclosed a very low level of replicability of these fit algorithms, which is clearly due to their tendency to become trapped inside a specific local maximum. Hereby, the "choice" of the local maximum depends on the random seed value used each time as the starting point for each analysis, which explains why each fit attempt leads to a completely different "best fit" result.

An unexperienced user, unaware of such seed effects, may conduct repeated analyses with these fit algorithms, using always the same, fixed internal seed values. Using always the same seed value will most likely lead to obtaining always the same output result, i.e., to VBGF parameters obtained from one specific local maximum, which may be far away from the overall best fit. This inherent pitfall of both fit optimisation methods may give an erroneous impression of extreme replicability of these methods and lead to overconfidence in their results.

*3.3 Full nonparametric bootstrap (FBoot experiments)*

In contrast to single-fit methods, the new bootstrap approach presented in this study (FBoot routines) proved to be very robust, replicable and accurate (Figs. 4b,d and 5b,d). In general, LFDs changed very little in their structure during reconstructing and resampling (Fig. 3). Especially the first, very strong size classes (i.e., the strong modes) composed of small-sized individuals were virtually unchanged during resampling and restructuring, showing a high degree of constancy. While the overall shape of the LFD distribution did not change considerably within the first two or three cohorts, the shape, location and size of the modes composed of large-sized individuals were completely different between each resampled LFD (Fig. 3), mainly due to their low numbers per mode. This is easily explained, since resampling with replacement from a strong mode (e.g. 1,000 ind.) would hardly produce any noticeable changes in mean size of this mode, but resampling from a minuscule cohort of large individuals (e.g. 20 indiv.) will most likely produce very variable results after each resampling run. Thus, the accuracy of the ELEFAN-based methods, as assessed by the current bootstrap experiments, is strongly affected by the data available for the weakest modes, generally the largest fish.

When overlaying bi-dimensional 95% confidence contours obtained with FBoot on their respective RSA plots (Fig. 6), it became evident that these bootstrap-derived contours have a clear relationship to the RSA plateau, being generally located within the plateau (Fig. 6). This shows a good consistency in the data and methods. FBoot contours were always smaller than their corresponding RSA plateaus, for the two datasets analysed in this study.

*3.4 Comparing PBoot and FBoot – assessing the relative magnitude of error sources*

For the *A. alba* data set, CIs obtained for growth parameters with FBoot were significantly and considerably larger than those obtained by PBoot experiments (Tables 3 to 5). This, very small *Abra alba* data set showed a low pseudo-$R^2_{Phi'}$, from to 54% to 68%, indicating the need for more samples and a larger LFD data matrix (Table 5). For a large, perfect synthetic data set, pseudo-$R^2_{Phi'}$ was very high (88 to 100%), indicating an excellent fit of the VBGF model.

Considering all three VBGF parameters (pseudo-$R^2_{Linf}$, pseudo-$R^2_K$, pseudo-$R^2_{Phi'}$), pseudo-$R^2$ values for the small *A. alba* dataset varied from 0.45 to 0.85 (Tables 3 to 5). This showed that for this dataset, a significant and relevant portion (15 to 55%), of the overall uncertainty was derived from sampling error, i.e., from the uncertainty or error derived from stochastic sampling from the population during fieldwork. In this particular case, sampling more individuals in the field (larger N, more months) would thus provide considerably more precise estimates, for $L_\infty$, K and Phi'. This was to be expected for such a small data set that contains only seven sampling months and only 675 individuals.

Conversely, for the much larger and perfectly shaped synthetic dataset synLFQ5b, CIs obtained with PBoot and FBoot were generally not significantly different, neither for $L_\infty$ nor for K, or CIs obtained from FBoot experiments were only slightly larger (Tables 3 to 5). This indicates that seed effects and other sampling-independent method artefacts were the clearly dominant error source for this dataset, and that the error due to sampling in the field was negligible or small, assuming a perfectly random representation of the population in these samples, which was true for this synthetic dataset. This shows that its sample size was sufficient for ELEFAN-based analyses, which is easily explained by the large sample size (12 months and 36,711 individuals) and clearly defined cohort peaks of this synthetic data set.

*3.5 Accuracy of bootstrapped length-based methods*

The approach proposed here correctly and reproducibly found the "true" initial parameters for the synthetic synlfq5b data, within 10% bias (for the natural *A. Alba* data, the "true" parameters are unknown). The target values of K (0.5 $y^{-1}$), $L_\infty$ (80 cm) and Phi' (3.505) were always well within the obtained CIs (Tables 3 to 5). Furthermore, in all scenarios, the median values of the posterior distributions were very close to the target values, within 10% for K, within only 4 % for $L_\infty$, and within only 0.4 % for Phi' (Tables 3 to 5), showing a very high level of accuracy of this bootstrapped length-based method, especially for Phi'.

1. **Discussion**

This study shows that modern fit algorithms can be used for bootstrapping. In contrast to current one-fit-only methods, this new approach displayed a high level of reproducibility and accuracy. Thus, this new, bootstrapped approach can be recommended for regular growth curve fitting. Also, it provides CIs for all parameter estimates, and a new index of data quality (pseudo-$R^2_{Phi'}$).

*4.1 Critical evaluation of RSA and K-Scan – is there a need for new methods?*

Hitherto, the two standard methods used the find the "optimum" VBGF curve fit have been RSA and K-Scan, as embedded in the FISAT II software (Gayanilo et al., 1987). These two methods have been used in many studies worldwide, especially for data-poor scenarios in developing countries. When comparing both methods, RSA has the big advantage of allowing for an unconstrained, simultaneous search for any possible combinations of K and $L_\infty$, while the K-scan method depends on the questionable underlying assumption that it is possible to obtain a single, reliable $L_\infty$ value from other methods and search for K only.

RSA is a robust, intuitive and graphical method, it has, however, several issues and limitations. For example, RSA does not allow for any assessment of seasonality in growth, since the seasonal growth parameters C and ts cannot be considered in a common two-dimensional RSA plot. In theory, it is actually possible to use RSA for the more complex soVBGF (seasonally oscillating VBGF), including plots for the seasonal parameters C (seasonal amplitude) and ts (summer point, as used in TropFishR) and thus build a set of bi-dimensional RSA plots (or a set of higher dimensional matrixes and plots), as originally suggested when the method was first presented, (Brey et al., 1988). However, this is not done in actual practice, probably due to the immense computation time that would be necessary for the calculation of all possible high-dimensional combinations and the construction of high-resolution, multidimensional RSA matrixes and plots, and due to the fact that no ready-to-use software exists for multidimensional RSA routines. Already the constructing of a relatively simple bi-dimensional RSA plot (Figs. 2 and 6) can be very time-consuming, depending on the resolution chosen and the size of the data set. Another limitation of the RSA method and all other currently available VBGF fit methods, is that no CIs can be provided for K and $L_\infty$, or any other output parameter estimates.

Another, often overlooked basic problem is that the output of the RSA plot is highly dependent on its input grid parameters, as shown in this study (compare Figs. 2 and 6). Multiple maxima are a common issue for RSA analyses (Isaac, 1990; Pauly and Greenberg, 2013; this study). Depending on the conformation of the grid, several maxima usually appear, where the number, position, and relative height of these maxima is very sensitive to subtle changes in the choice of grid points (this study). Thus, the overall optimum (the combination of K and $L_\infty$ that produces the highest peak) obtained by RSA may be very subjective. The choice of the "best" local maximum is thus dependent on the subjective *a priori* definition of the RSA grid, although subsequent RSA explorations may focus searches on a more refined area with smaller parameter increments.

A simplistic, intuitive approach would be to consider the widely cited RSA "banana-shaped" plateau (Pauly and Greenberg, 2013) as a proxy of a 95% confidence contour. However, this method does not provide any CIs, confidence contours nor any parameters for seasonal growth. Yet, this study showed that there was a clearly visible relationship between the RSA plateau and FBoot 95% confidence contours (Fig. 6). The FBoot contours were generally located well inside the plateau, which indicates a good consistency of these methods. However, a direct numerical or graphical estimation of CIs from RSA plots is still not possible, other than stating that 95% confidence contours are most likely somewhere within a subjectively defined plateau. There is still no procedure available, or to expect, that extracts quantitative information regarding uncertainty from a RSA plot, without proper bootstrapping.

*4.2 Reproducibility of the fit methods - beware of local maxima*

As expected, the full bootstrap routine (FBoot) always produced different results within each run, since it analyses a different set of resampled LFD data in each run. Most surprisingly, the partial bootstrap routine (using always the same LFD data and the same settings) also yielded a completely different result for each run, depending on the seed values used. Clearly, both tested fit algorithms (ELEFAN_GA and ELEFAN_SA) get trapped in specific local maxima, a pitfall that has already been identified since the earliest days of ELEFAN (Shepherd, 1987; Rosenberg and Beddington, 1987), but had not yet been appropriately addressed and solved. In a recent analysis of perfect synthetic data, Pauly and Greenberg, (2013) showed that these maxima can be plentiful, of very similar height and distributed along the characteristic "banana-shaped plateau". The best and most widely recommended approach has been to plot a K-scan (while fixing $L_\infty$ with dubious external methods) or a RSA plot and then to look for the most prominent maxima. Acknowledging this central issue, most ELEFAN handbooks and tutorials recommend, as general rule, that the user should always be "analyzing several local maxima" (Mildenberger, 2017), but there has not yet been any advice or rule available to choose between these multiple maxima. More seriously, there had been no published study of the effect of subjective choice of grid settings on the location and relative height of Rn peaks, and thus on the outcome of the analysis, which is probably its most serious pitfall. The huge amount of population studies and stock assessments based on such analyses may have had a substantial amount of previously unreported subjectivity, relying the researchers' "intuition".

Clearly, optimization-based fit functions, such as ELEFAN_GA or ELEFAN_SA are a big leap forward from these earlier approaches. However, here the user always will have to choose one seed value for starting the search for the overall optimum. When doing repeated analyses, the result may be always the same, since using the same seed for the fit usually leads to the same local maximum. This may give the user the erroneous impression, that these single–fit methods are very robust and 100 percent reproducible in their results, when in fact, they are not, as shown in PBoot experiments. Similar to the subjectivity issue observed for RSA, this seed effect is probably the single most dangerous pitfall in ELEFAN-based fit algorithms ELEFAN_GA and ELEFAN_SA. One simple, but generally neglected way to overcome this issue is to run the fit algorithms many times, explicitly choosing different seed values (as in the PBoot experiments), or by full bootstrapping.

*4.3 The importance of not fixing $L_\infty$*

Reducing subjectivity of older paper-and-pencil methods was the main original reason for creating the first electronic length-frequency analysis tools (Pauly and David, 1981; Pauly, 1986). However, issues related to uncertainty and subjectivity in LFD analyses have since then been generally neglected. It seems that most authors simply have assumed that the ELEFAN I method works fine, ignoring any problems related to uncertainty and accuracy. This is probably due to erroneous confidence in *a priori* fixing $L_\infty$ (by using $L_{max}$ or by the Wetherall plot method), a technique that has been widely used to miraculously "solve" all problems related to uncertainty and accuracy in past analyses (Schwamborn, 2018). All tutorials suggest ascertaining $L_\infty$ *a priori* by these methods, before starting a curve fit with ELEFAN I (e.g., Sparre and Venema, 1998). Not surprisingly, this "fixed-$L_\infty$" approach has been generally accepted and used in the past decades to assess the growth of many fish and invertebrate populations worldwide. Yet, recent simulation experiments (Schwamborn, 2018) have proven that the Wetherall plot method is not suitable for fixing $L_\infty$, and that there is no relationship whatsoever to be expected *a priori* between $L_{max}$ and $L_\infty$. Thus, there is no way to determine $L_\infty$ *a priori*, independently of the overall VBGF model for any given population. Using a fixed, unique $L_\infty$ value (determined by such dubious methods) during ELEFAN I analyses will always lead to extremely constrained results for K, Phi', etc., and to extreme overconfidence in such potentially biased results (Schwamborn, 2018).

Even more seriously, *a priori* fixing $L_\infty$ may lead to severe systematic bias (Schwamborn, 2018). A common phenomenon in natural populations under severe fishing pressure is the absence of large individuals, due to gear avoidance or overfishing (Pauly et al., 1998). This will lead to a significant underestimation of $L_\infty$, when using the $L_{max}$ approach and when fixing $L_\infty$ with the Wetherall plot method, which also biased towards $L_{max}$ (Schwamborn, 2018). Then, due to the well-described interaction between the growth parameters K and $L_\infty$, this invariably leads to an overestimation of K. Finally, an overestimated K leads to an underestimation of a population's vulnerability to overfishing. This hitherto unnoticed, potently vicious systematic bias may lead to overoptimistic fisheries management practice and to the destructive overexploitation of vulnerable and severely threatened populations (Schwamborn and Moraes-Costa, subm.). The bootstrap-based approach proposed in this study is a promising path towards overcoming this generally neglected issue, allowing for a simultaneous, unconstrained search of all parameters, including $L_\infty$, as a future standard approach.

*4.3 Comparing the fit methods and settings - speed vs precision*

Seasonality in growth is a generally acknowledged fact, which includes most tropical populations (Longhurst and Pauly, 1987). In contrast to RSA, the new fit methods ELEFAN_SA and ELEFAN_GA (and their bootstrapped versions) both permit the estimation of seasonality in growth and are fast enough for routine use, as shown in this study. Additionally to providing estimates of seasonality parameters C and ts (Mildenberger et al., 2017), the bootstrapped versions (ELEFAN_SAboot and ELEFAN_GAboot) provide CIs for both seasonality parameters, although including seasonality does increase computation time considerably. Therefore, in many cases, compromises between fit accuracy (fit algorithm settings) and the consideration of seasonality will have to made, as long as computation time is limiting. In spite of a considerably reduced speed, seasonality in growth can be explicitly modelled and quantified, using the above mentioned algorithms and their bootstrapped versions. At present, many users with limited computation power and time will still have to accept a compromise between speed and accuracy.

Since ELEFAN_GA can use several cores simultaneously, and multi-core processors are becoming increasingly available at lower cost, a bootstrapped version of ELEFAN_GA will become much faster for common users in the very near future. For illustration, the outputs (the best fits, the median of best fits, confidence envelope or confidence ellipse) could then be plotted using a high-resolution RSA heatmap as a background (e.g., as in Fig. 6).

Several fundamental aspects of LFD analyses, such as and seed effects and local maxima have been addressed in this study. However, still many other potential issues of subjectivity and potential artefacts remain to be investigated, such as the effects of MA span, class width and the choice of each specific fit setting within the complex optimization algorithms.

Early enthusiasm over length-based methods led to a huge number of studies with LFD analyses, often based on extremely small datasets. The large CIs observed in the present study suggest that much larger datasets may be necessary to obtain informative growth estimates than previously imagined. A very limited sampling effort, such as for *A. alba*, is clearly not sufficient for confidently estimating growth parameters from LFD data, considering the vast CIs ascertained in this study. Future sampling efforts will probably have to focus on sampling very large numbers of individuals regularly over a long time period, and a establish minimum number of monthly large individuals, as to be able to provide narrow, useful CIs for growth parameters, that can be used for subsequent population and stock assessment models, considering uncertainties and risks.

This study is a first step into unveiling the uncertainties and limitations inherent of length-based methods, opening up a new area of investigation and indicating a new path towards more reliable and accurate approaches. In the near future, increased computing speed at lower costs will enable many more users to perform such thorough analyses on a routine basis.

*4.5 The importance of large individuals*

The present experiments showed that uncertainty in growth estimates is strongly influenced by the reliable, regular and sufficient sampling of large-sized individuals. In general, the strength of the weakest cohorts (i.e., large individuals) defines the accuracy of the method, not the overall N.

Capturing a few more large individuals every month can thus have stronger positive effect on the accuracy of growth estimates than hundreds of small individuals.

Maximum effort should thus be invested during the field sampling activities to obtain a reasonable number and equal and best possible repartition of individuals, covering the full length spectrum (Schwamborn, 2018). The biologically and historically possible length spectrum may be looked up from online databases, such as FishBase (Froese and Pauly, 2000), prior to field sampling, as an aid in advising these efforts. In many cases, however, large individuals will simply not be available anymore in the ecosystem, e.g., due to overfishing and climate change (Brander, 2007; Cheung et al., 2013).

*4.6 Usefulness of the new pseudo-R² index – quality data for optimized models*

In this study, a new index for LFD data and growth model quality (pseudo-$R^2$) is presented, based on the ratio of matrix-information-independent error ($CI_{PBoot}$) to the overall error of the ELEFAN procedure ($CI_{FBoot}$). This index is directly related to the information content in the original LFD data, low pseudo-$R^2$ meaning "poor" data. This new index has a variety of simple upcoming practical applications. For scenarios with low pseudo-$R^2$, as in the *A. alba* example, with pseudo-$R^2_{Phi'}$ of only 0.54 to 0.68, where only 54% to 68% of the total uncertainty in Phi' were unrelated to matrix information content (matrix size and VBGF-like shape), up to 32 to 46 % of the total error in Phi' was due to its matrix information content and thus directly related to sample size (N individuals), that determines the relevant LFD matrix size (number of sampling months vs number of size bins). A larger sample size (N individuals) will generally allow for smaller bin widths

(higher resolution) and thus for a lager LFD matrix, with more size bins. For a scenario with high sample-size-related error, as in the *A. alba* dataset, the researcher can conclude that going back to the field and obtaining more samples (i.e., adding sampling months and size bins) would be a promising strategy for increased precision in estimates of growth.

Conversely, the case of the perfectly shaped and very large synthetic synlfq5b dataset, its high pseudo-R² (up to 1), indicates that all effort can be concentrated in the optimization of the fit algorithms (choosing slower, more precise settings and dedicating more time into bootstrap analyses), as to optimize and narrow down the CIs of growth parameters.

These observations confirm that pseudo-$R^2_{Phi'}$ values close to 1 indicate highly informative LFD data with excellent model fits. Growth models with low pseudo-$R^2_{Phi'}$ indicate a poor adjustment of the overall VBGF model to the LFD data and a dominance of non-VBGF error in the data. Pseudo-R² is related to sample size, overall matrix size and to the alignment along a VBGF curve *vs* overall error. The pseudo-R² index is conceptually analogous to the coefficient of determination R² used for linear models. Furthermore, the basic setup of the pseudo-R² index (pseudo-R² = "matrix-information-independent error" / "total error") may be considered to be analogous to R² (R² = "explained variability" / "total variability") . Thus, pseudo-$R^2_{Phi'}$, is a promising new index of model goodness-of-fit and data quality, that condenses relevant information regarding the relative magnitude of determination and stochastic error in such complex data and models into one single index. Similar to common R², this new index (pseudo-$R^2_{Phi'}$) is also determined by sample size and information content (i.e., alignment according to the model) of the data (Zar, 1990).

Other forms of pseudo-R² have been proposed, for other non-trivial applications, such a logistic regression (Cox and Snell, 1989; Magee, 1990). One striking difference between the pseudo-R² and common R² is that pseudo-R² may show values equal to 1.0 (Table 5) for perfect synthetic LFD data, even when there is clearly a considerable amount of non-explained variability present in these data. Conversely, a common linear model with "R² = 1" means that 100% of all variability is explained by the model, which means zero error, a phenomenon unlikely to be observed in any ecological data. Similarly, a 100% perfect alignment of, modes, as in the synthetic synlfq5b dataset, is equally unlikely to be observed in any samples taken from natural populations. Still, one possible pitfall of this new index is that pseudo-R² values close or equal to 1 may also be observed when the fit method used is very imprecise (very large $CI_{Pboot}$). So, care has to be taken to calculate pseudo-R² only for precision-optimized bootstrap routines, using exactly the same fit parameter settings and seed value distributions for both PBoot and FBoot, as done in the present study.

The pseudo-R², additionally to its academic interest for statisticians, can be a useful tool for the design and improvement of sampling and analysis strategies in hands-on population studies. Field sampling efforts and high-precision bootstrap experiments can both be very time-consuming and costly. This new, simple and intuitive index may be useful in the evaluation of the quality of LFD data and in the choice between possible strategies for the improvement of model reliability and risk management.

*4.7 A plea for new precautionary approaches, including all reasonable sources of uncertainty*

The ELEFAN I approach is affected by many sources of inherent uncertainty and subjective choice, such as the unavoidable stochasticity of sampling (this study), natural variability in growth (Isaac, 1990, Schwamborn, 2018), gear selection effects (Schwamborn, 2018), occasional occurrence of few large individuals (Schwamborn 2018, this study), bin width (Pauly , 1984; Hoenig et al., 1987; Sparre, 1989; Wolff, 1989; Isaac, 1990), MA span (Pauly, 2013; Taylor and Mildenberger, 2017), RSA grid design (Pauly and Greenberg, 2013, this study) and seed effects (this study). Considering all these sources of uncertainty and the complex non-linear interactions, precautionary approaches should be considered for future analyses. Future studies will consider seasonality in growth and probably test a range of MA values and bin widths, within a range of settings, and will certainly require a considerable amount of computation time. The experiments presented here are a starting point for this new research steam of thorough assessments of uncertainty in length-based methods.

For more thorough analyses, the combination of length-based analyses with ageing or marc-recapture should be recommended, especially if a more accurate estimate of growth parameters is needed. However, there are few tools available for proper merging of length-at age data and LFDs, considering uncertainty and variability, and selective mortality, and mark-recapture (Schwamborn and Moraes-Costa, subm.). Standard tools that allow the combination of external results (e.g., from reading of hard structures) and LFD analysis, that include their inherent uncertainty are still to be developed.

The comparison of partial and full bootstrap experiments and their outputs (CIs and confidence contour plots) provided a first glimpse into the reliability and accuracy of modern ELEFAN-based fit methods. Yet, the approach shown in this study is not simply an experiment of merely academic interest and merit, used for the assessment of uncertainty. It is a new robust and useful method to obtain key data to be used in population studies and stock assessments. This method has the advantage that it avoids the dreadful pitfalls of seed effects and local maxima. Even more importantly, applying a completely unconstrained search for $L_\infty$ and K avoids the serious systematic errors (e.g. overestimation of K) that can happen when artificially constraining $L_\infty$ *a priori* (e.g., from the largest individuals or by non-reliable methods, such as the Wetherall plot (Schwamborn, 2018). Also, it will assist in avoiding overconfidence in such estimates and subsequent population and stock assessment procedures, and can be the base for a completely new suite of approaches and tools for robust risk assessment in marine resource management.

*4.8 Consequences for fisheries management – know your risks*

Given the large list of well-documented collapsed or collapsing stocks (e.g., Hilborn and Ovando, 2014; García-Carreras et al., 2016), the decreasing overall biomass (Christensen et al., 2014) and trophic level (Pauly et al., 1998; Pauly and Palomares, 2005; Shannon et al., 2014) of populations under severe fishing pressure, there are emerging questions whether current management approaches have failed (Wolff, 2015; García-Carreras et al., 2016). Due to these historic and recent failures, changes in fisheries policy have been suggested, such as more precautionary approaches (e.g., Hilborn and Peterman, 1996) and such drastic measures as the overall prohibitions of any fisheries in large sections of oceans and coasts (Leenhardt et al., 2013; Wolff, 2015).

There are numerous compilations and descriptions of widely used standard approaches and models for stock assessment and resource management (e.g., Sparre, et al. 1989; Sparre and Venema, 1998; Hilborn and Walters, 1992; Lleonart, 2002), and a steadily increasing number of recent developments and new tools (e,g, Maunder et al., 2016; Mildenberger at al., 2017). One emerging line of thought, or new sub-discipline, has been the assessment of the risk of overfishing and stock collapse, usually using Mote Carlo simulations, jackknifing and bootstrapping (Caddy and Defeo, 1996; Hilborn and Peterman, 1996; Francis and Shotton, 1997; Seijo et al., 1998; Seijo et al., 2004; Magnusson et al., 2013).

Usually, each risk assessment study uses a discrete, unique list of sources of uncertainty. Hilborn and Peterman (1996) considered the uncertainty (1) in the estimates of fish abundance (2) in the structure of the mathematical model of the fishery; (3) when estimating model parameters; (4) in future environmental conditions; (5) in the response of users to regulations; (6) in future management objectives; and (7) in economic, political and social conditions (see Hilborn & Peterman (1996 for details). Caddy and Mahon (1995) included other sources of uncertainty, such as (1) variability introduced by the effects of variable abiotic factors (2) effects of ecological interdependencies; (3) fluctuations in costs and product prices; (4) variations in fishing effort (5) variability in the behavior of policy makers.

However, little has been done in any of such assessments to investigate the uncertainty inherent to the very foundations of any such model: growth estimates that are used to calculate mortality (e.g., by using the length-converted catch curve method, Pauly, 1984) and for all subsequent analyses. Instead, it is generally assumed that growth estimates are perfect and subject to zero or very low uncertainty. Thus, it is very likely that previously and currently applied risk assessments and other models have drastically underestimated the overall uncertainty and risk of fisheries management decisions, especially (but not only) in cases when growth parameters were derived from poor-quality LFD data.

There are five direct benefits of the proposed `ELEFAN_Boot` approach: (1) unconstrained search, (2) better reproducibility and accuracy, (3) assessment of uncertainty (CIs) inherent of all VBGF parameter estimates, including seasonality, (4) assessment of the information content, or quality of LFD data (e.g. by comparing CIs obtained with PBoot vs FBoot in the pseudo-$R^2$), and (5) possibility to export these CIs (or, better, the raw posterior distribution data) to subsequent analyses, such as mortality calculations, stock assessments, simulations and bioeconomics or socio-ecological risk assessment studies.

*4.9 Conclusions and Outlook*

This study demonstrated a promising path towards a robust and reliable approach to LFD analysis. The new approach presented here was incorporated into a set of ready-to-use R functions and is the basis for the new R package 'fishboot' (Schwamborn et al., 2018a). This new package will most likely soon become standard for stock assessments, especially under data-poor scenarios.

In the near future, researchers will be unable to imagine living in a world where growth parameters were estimated as "one result only", without any CIs, without pseudo-$R^2$ and "p" values, where there were different results obtained from every fit attempt, even when using the same method. This is how we remember the old days when common linear regression outputs were drawn on a sheet of paper as a single straight line, without any confidence envelopes for the slope or any further tests and indices. Clearly, we are stepping into a new era of robust length-based methods.

**Acknowledgments**

Many thanks to Daniel Pauly for initiating this endeavour many years ago and for numerous inspiring comments. Thanks to M. L. 'Deng' Palomares for encouraging the formation of a new working group on length-based methods. Many thanks to Matthias Wolff for co-organizing a successful and productive workshop on length-based methods at ZMT Bremen in May 2018 and for his enthusiasm regarding this new project. RS received a productivity fellowship from the Brazilian National Research Council (CNPq). Many thanks also to Rainer Froese and to all participants of the workshop for helpful comments.

**FIGURES WITH CAPTIONS**

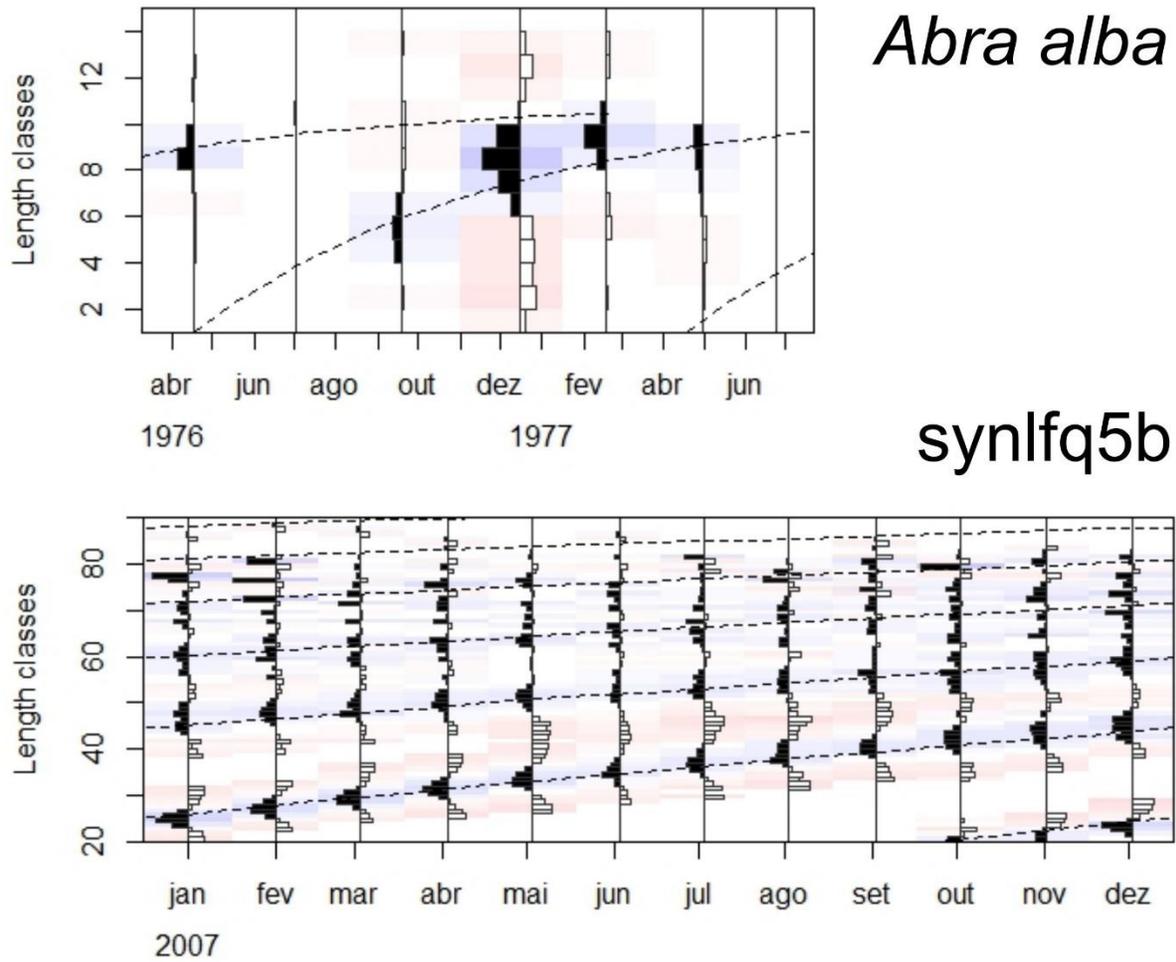

**Fig. 1: Length-frequency distributions for the *Abra alba* and synlfq5b data sets.** The horizontal bars show the differences between moving averages and actual class strength. Black bars indicate positive values (peaks), white bars indicate negative values (troughs). Moving average spans are 9 and 11 for *Abra alba* and synlfq5b, respectively. VBGF curves shown are $L_\infty$ = 9.98 mm, K = 2.96 $y^{-1}$ for *Abra alba* and $L_\infty$ = 105.6 cm, K = 0.28 $y^{-1}$ for synlfq5b.

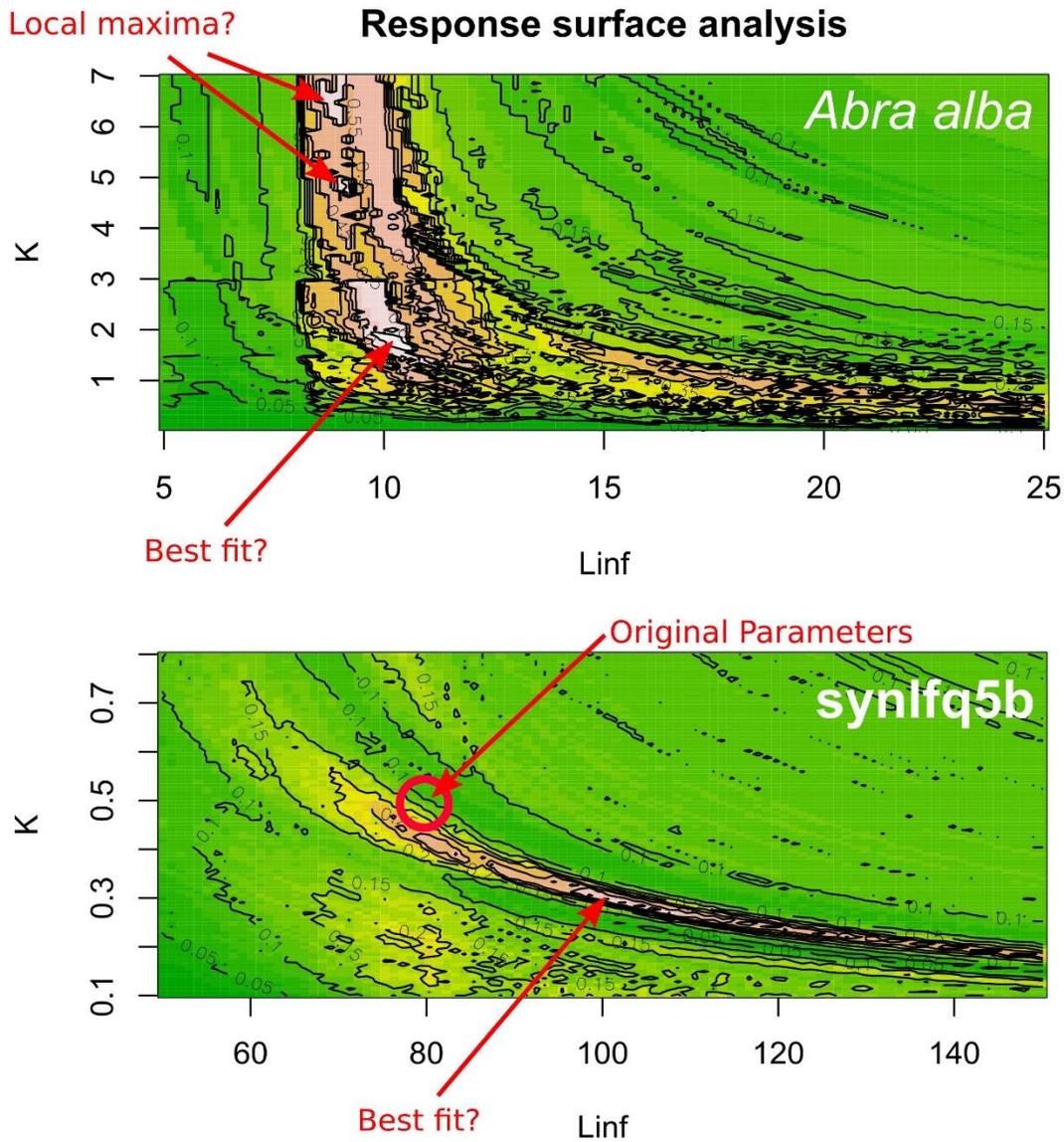

**Fig. 2: Response surface analysis (RSA) plots for the *Abra alba* and synlfqb data sets.** Gridding method: "optimise" (default), resolution: 100 x 100. Best fit VBGF curves indicated are $L_\infty$ = 9.98 mm, K = 2.96 $y^{-1}$ for *Abra alba*, and $L_\infty$ = 105. 6 cm, K = 0.28 $y^{-1}$ for synlfq5b. Note that for synlfq5b, the original parameters, that were used to build the data, were $L_\infty$ = 80 cm and K = 0.5 $y^{-1}$ (red circle). Best fit: highest Rn value.

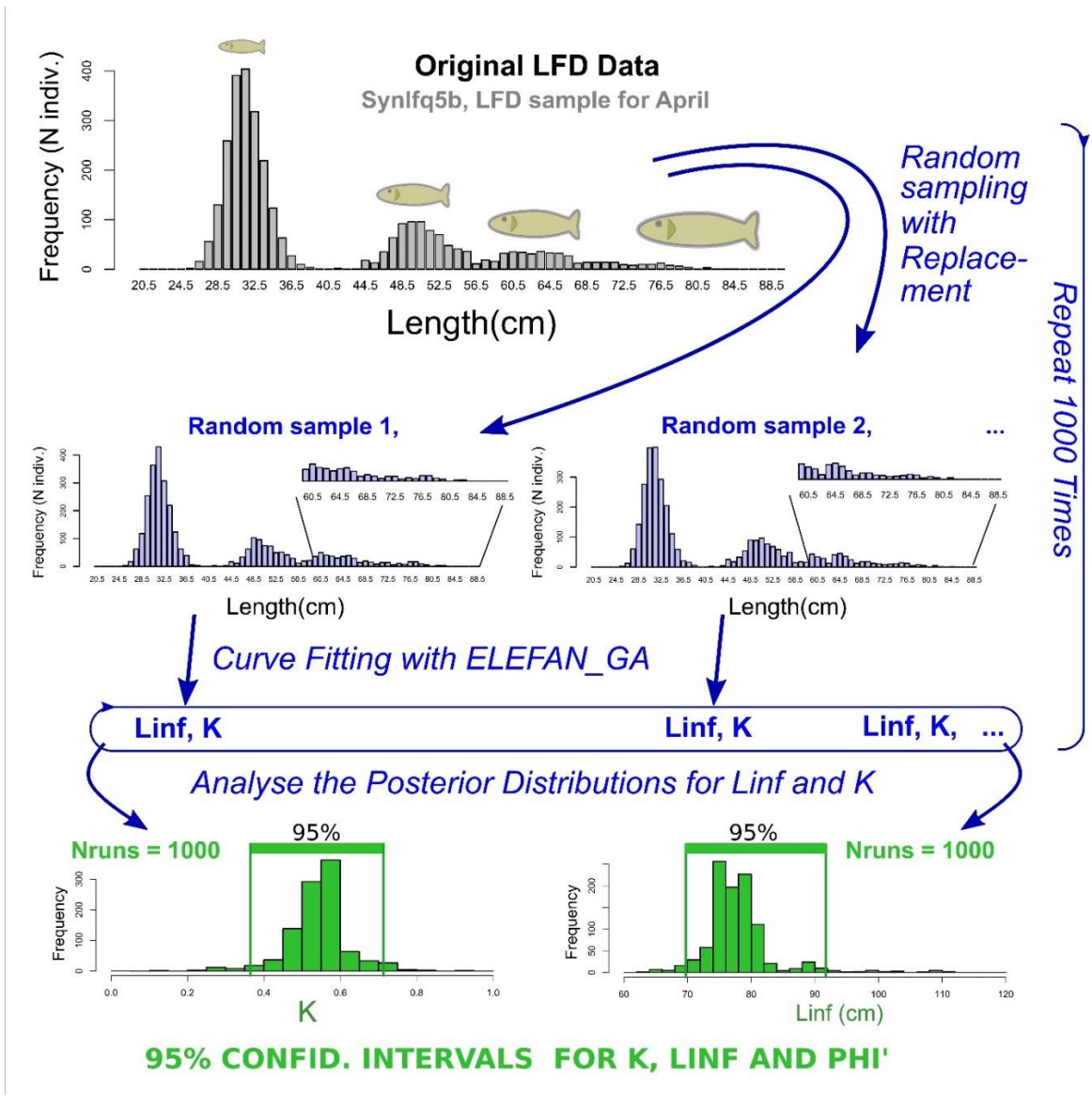

**Fig. 3: Flow diagram of the full bootstrap routine (FBoot experiments).** Note the differences in shape between the two resampled LFDs, especially within the largest size classes (L > 70 mm).

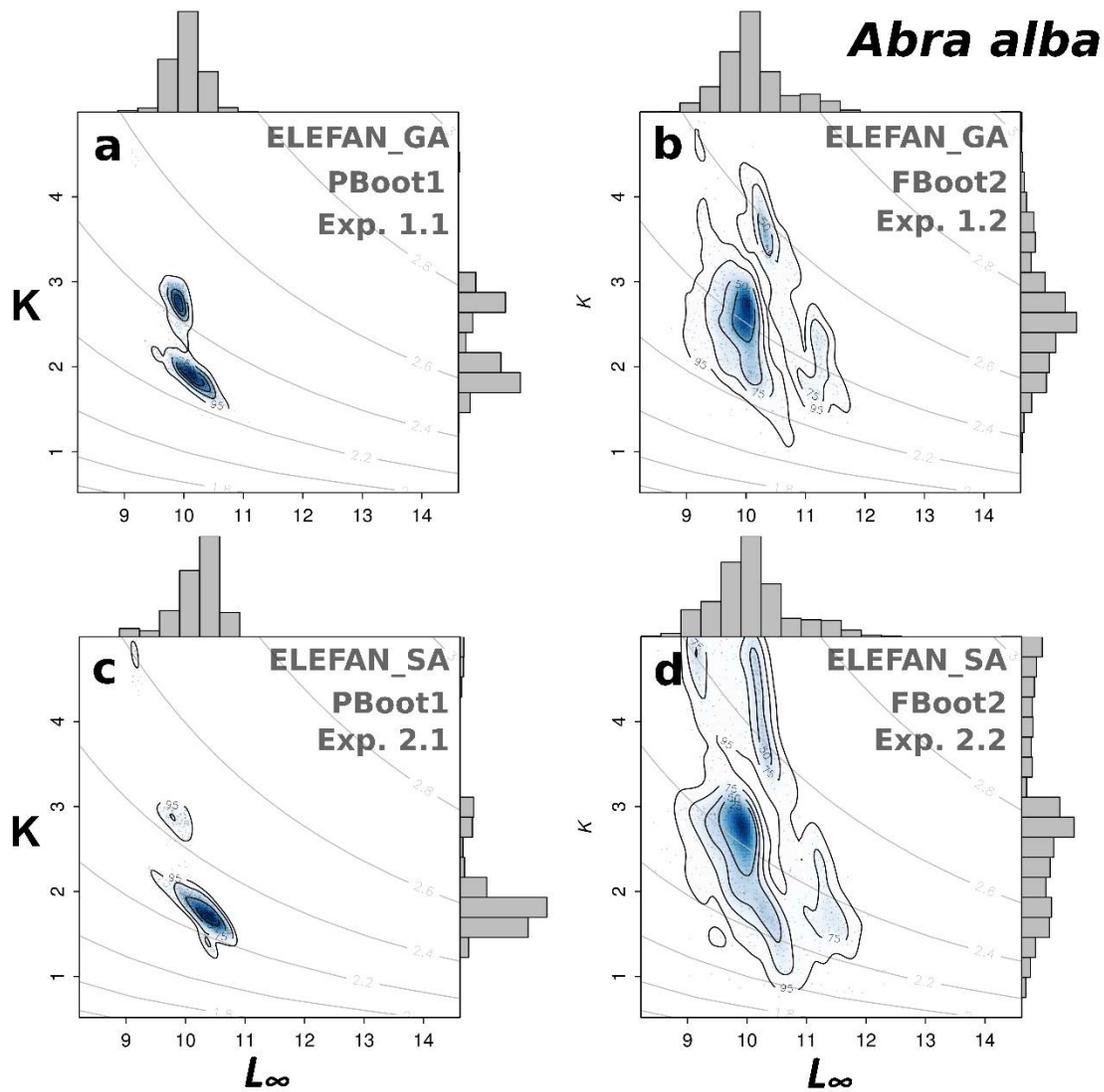

**Fig. 4: Results of four bootstrap experiments (Exp 1.1. to Exp. 2.2) using the *Abra alba* data set.** Each graph is a K *vs* L∞ scatterplot with 95% percentile contour (bivariate kernel distribution). Each blue dot represents one result obtained by optimized fit. Nruns = 1,000 for each graph. Maxiter = 50, MA = 7. Grey lines: Phi' isopleths.

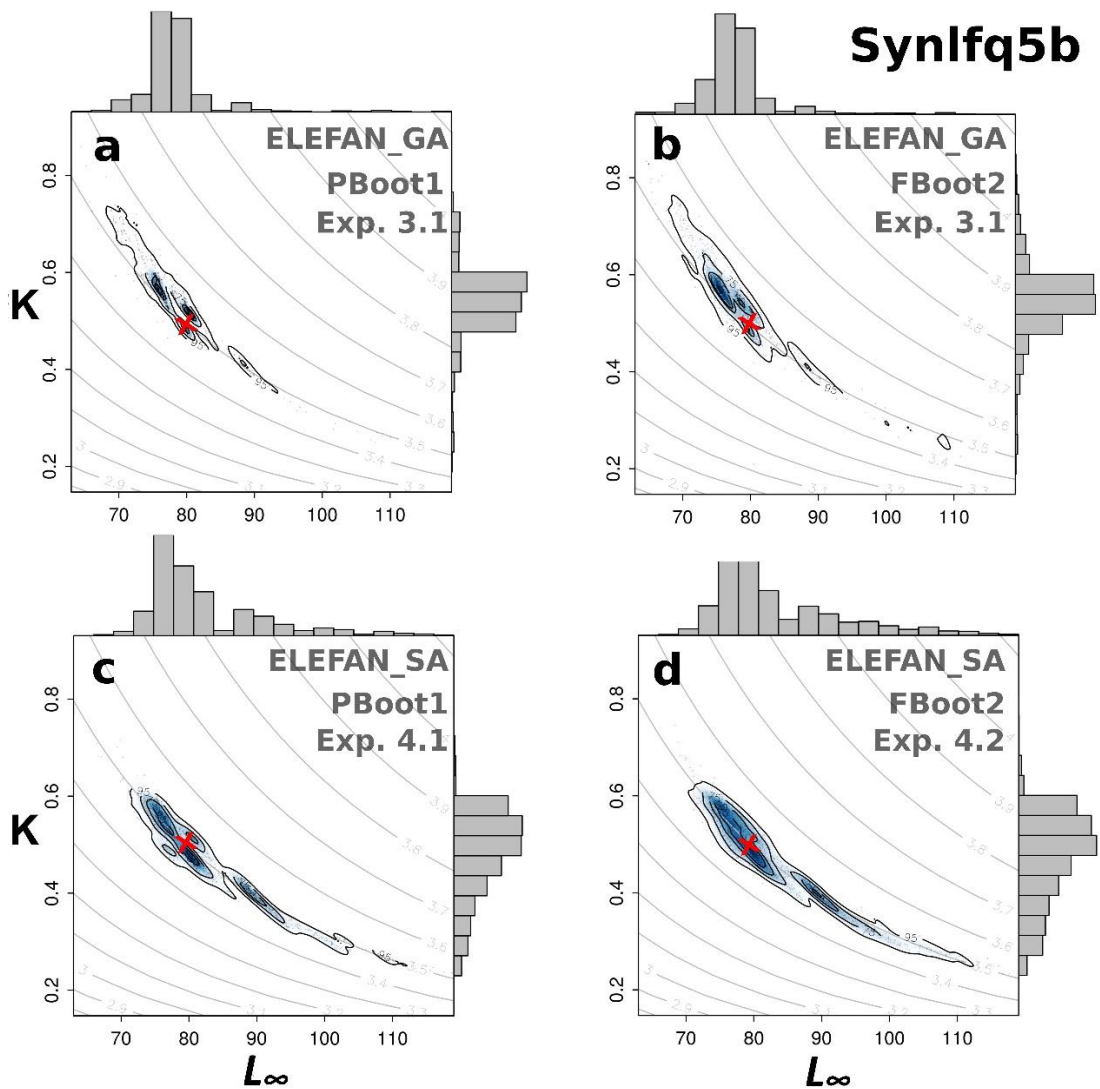

**Fig. 5: Results of four bootstrap experiments (Exp 3.1. to Exp. 4.2) using the *Synlfq5b* data set.** Each graph is a K vs $L_\infty$ scatterplot with 95% percentile contour (bivariate kernel distribution). Each dot represents one result obtained by optimized fit. Nruns = 1,000 for each graph. maxiter = 50, run = 10, popsize = 60, MA = 11. Grey lines: Phi' isopleths. Red "X": Original, "true" underlying parameters, that were used to construct the Synlfq5b data set ($L_\infty$ = 80, K = 0.5).

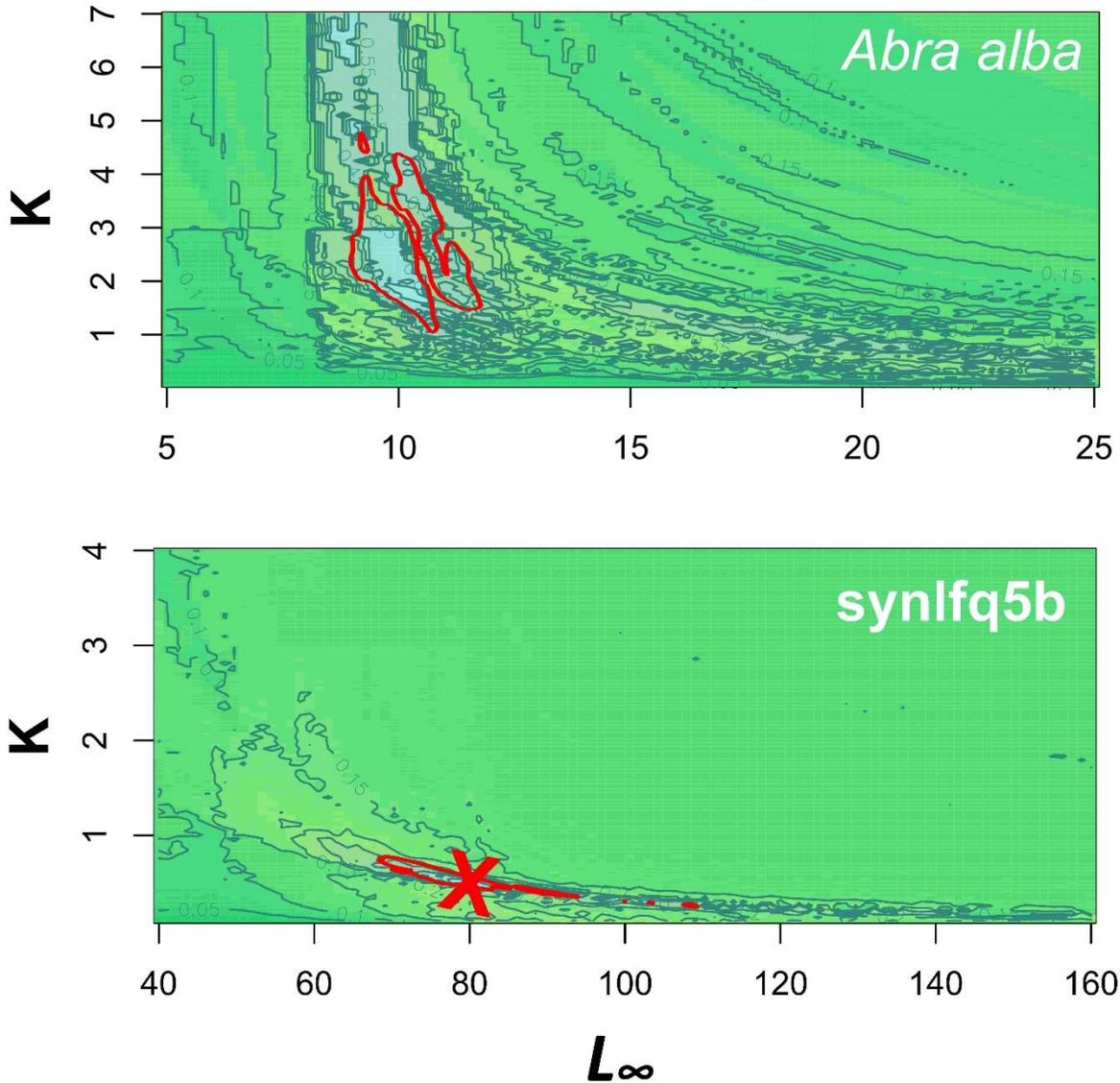

**Fig. 6: Response surface analysis (RSA) plots with overlaid 95% confidence contour for the *Abra alba* and synlfq5b datasets.** Red contours: kernel distribution contours of the 95% confidence space. Red cross: original "true" parameters (L∞ = 80, K = 0.5) used to build the synlfq5b data. Full Bootstrap, Nruns = 1,000, ELEFAN_GA fit algorithm optimized for precision with maxiter = 50. RSA resolution: 100 x 100 K *vs* L∞ values.

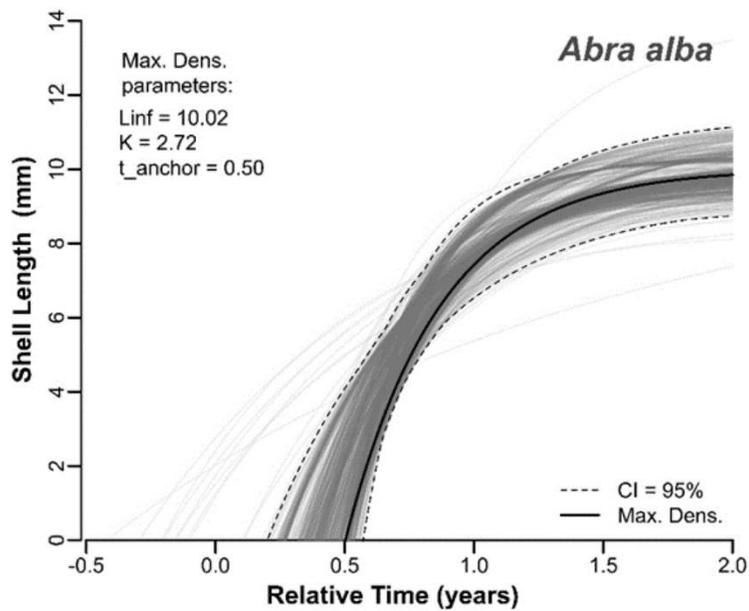

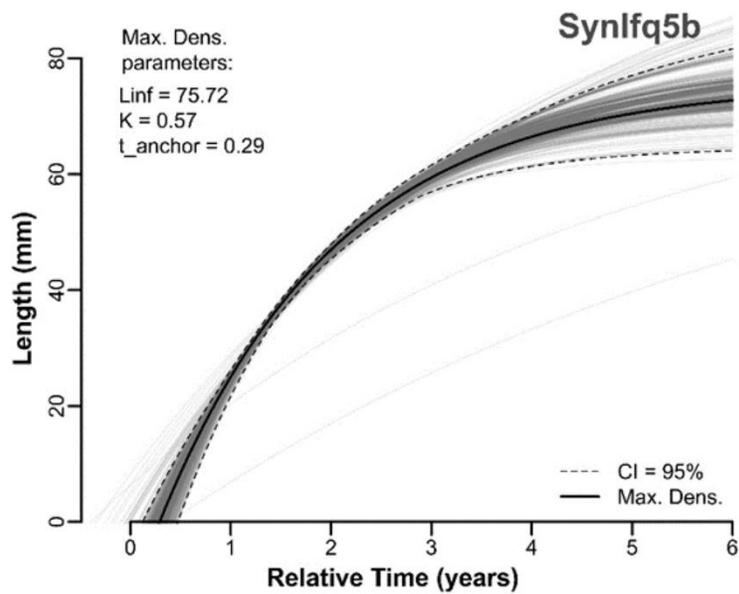

**Fig. 7: Curve swarms (grey lines) and 95% confidence contours (dashed lines) for the *Abra alba* and synlfq5b datasets.** Thick black line: Growth curve that represents the mode of the kernel density distribution (maximum density peak). Full Bootstrap, Nruns = 1,000. The ELEFAN_GA fit algorithm was optimized for precision (for details, see text).

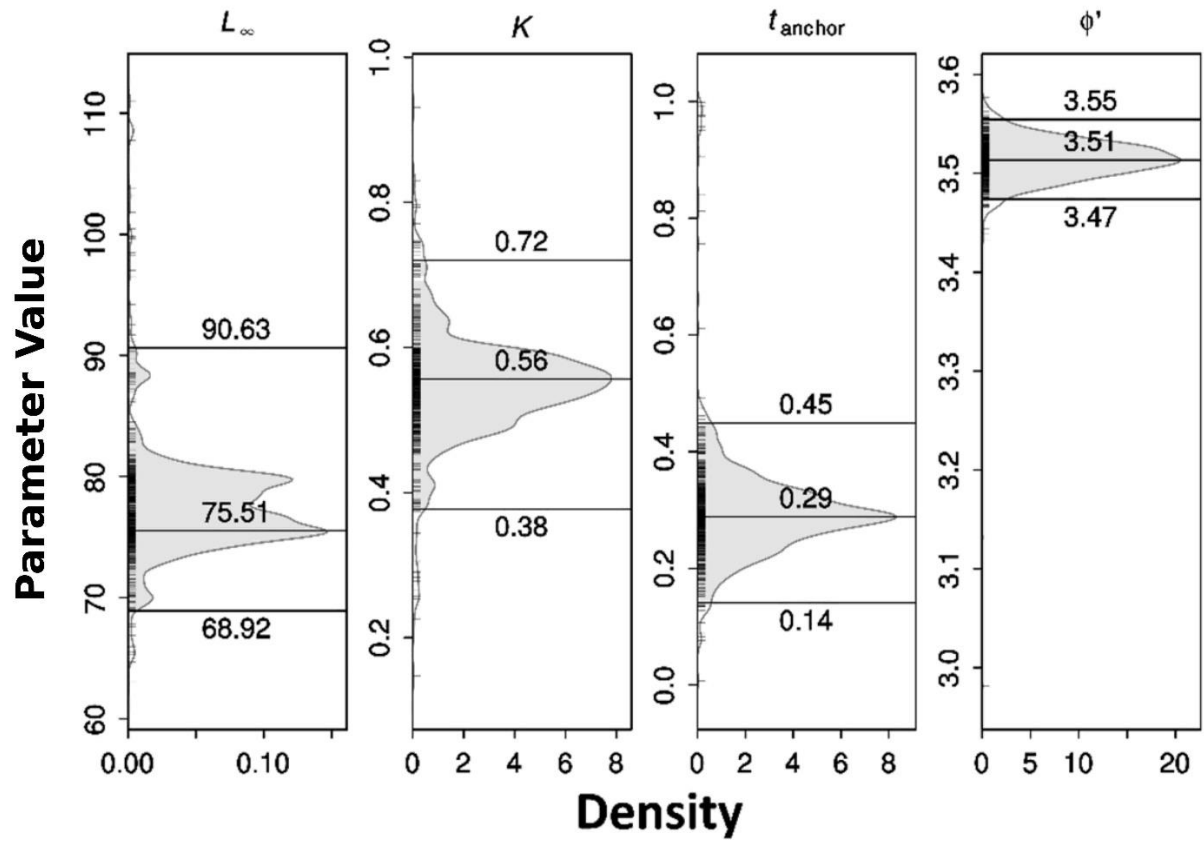

**Fig. 8: Example of quantitative outputs (95% CIs) obtained using the ELEFAN_GA_Boot method, optimized for precision (experiment 3.1 , Nruns = 1000, maxiter = 50, run = 10, popsize = 100, pmutation = 0.2, data: *synlfq5b*).** Kernel density distribution plots and univariate 95 % CIs of VBGF parameter estimates. Horizontal lines: location of the mode (maximum density peak) and of the lower and upper 95% quantiles of the univariate kernel density distributions.

# TABLES

**Table 1. Settings used for the fit algorithms.**

| Fit algorithm / argument | Search Settings* (*A. alba* / synlfq5b) |
|---|---|
| **ELEFAN_GA** | |
|   popSize | 100/60 |
|   pmutation | 0.2 |
|   maxiter | 50 |
|   run | 10 |
| **ELEFAN_GA_boot** | |
|   Bootstrap Runs | 1000 |
|   Bootstr. Time (min.) | 31*/80* |
| **ELEFAN_SA** | |
|   SA_temp | 5e+5 |
|   SA_time (sec) | 240 |
|   maxit | 500 |
| **ELEFAN_SA_boot** | |
|   Bootstrap Runs | 1000 |
|   Bootstr. Time (min.) | 47*/143* |

*: on a server with a 16-core processor.

**Table 2: Summary of the eight bootstrap experiments (Exp 1.1 to Exp 4.2).**

| *ABRA ALBA* | | |
|---|---|---|
| **FIT ALGORITHM** | **PBoot** | **FBoot** |
| **ELEFAN_GA** | Exp 1.1 | Exp 1.2 |
| **ELEFAN_SA** | Exp 2.1 | Exp 2.2 |
| *SYNLFQ5B* | | |
| **FIT ALGORITHM** | **PBoot** | **FBoot** |
| **ELEFAN_GA** | Exp 3.1 | Exp 3.2 |
| **ELEFAN_SA** | Exp 4.1 | Exp 4.2 |

**Table 3: Results for asymptotic length (L∞) from eight bootstrap experiments (Exp 1.1. to Exp. 4.2) using the *Abra alba* and synlfq5b data sets.** N = 1,000 runs, search settings were optimized for precision (e.g., maxiter = 50, run = 10, popsize = 100, pmutation = 0.2 for ELEFAN_GA and SA_temp = 5e+05, maxit = 500 for ELEFAN_SA). Median, lower and upper 95% CIs, 95% confidence interval widths are given. The last column shows the pseudo-$R^2$ ($pR^2$). n.s.: not significant., *: $p < 0.05$ ,**: $p < 0.0001$ (Quantile test).

| Exp. | Description | Median $L_\infty$ (mm) | lower $L_\infty$ (mm) | upper $L_\infty$ (mm) | conf. int. width $L_\infty$ (mm) | $pR^2$ ($CIW_{PBoot}/CIW_{FBoot}$) |
|---|---|---|---|---|---|---|
| 1.1 | alba, ELEFAN_GA, PBoot | 10.0 | 9.6 | 10.5 | 1.0 | |
| 1.2 | alba, ELEFAN_GA, FBoot | 10.0 | 9.2 | 11.3 | 2.2 | 0.45** |
| 2.1 | alba, ELEFAN_SA, PBoot | 10.3 | 9.2 | 10.7 | 1.5 | |
| 2.2 | alba, ELEFAN_SA, FBoot | 10.0 | 9.0 | 11.6 | 2.6 | 0.58** |
| 3.1 | synLFQ5b, ELEFAN_GA, PBoot | 77.4 | 70.3 | 91.0 | 20.7 | |
| 3.2 | synLFQ5b, ELEFAN_GA, FBoot | 77.0 | 69.7 | 91.7 | 22.0 | 0.94(n.s.) |
| 4.1 | synLFQ5b, ELEFAN_SA, PBoot | 79.9 | 73.4 | 107.8 | 34.4 | |
| 4.2 | synLFQ5b, ELEFAN_SA, FBoot | 80.0 | 72.6 | 108.1 | 35.5 | 0.97(n.s.) |

**Table 4: Results for the growth constant (K) from eight bootstrap experiments (Exp 1.1. to Exp. 4.2) using the *Abra alba* and synlfq5b data sets.** N = 1,000 runs, search settings were optimized for precision (e.g., maxiter = 50, run = 10, popsize = 100, pmutation = 0.2 for ELEFAN_GA and SA_temp = 5e+05, maxit = 500 for ELEFAN_SA). Median, lower and upper 95% CIs, 95% confidence interval widths are given. The last column shows the pseudo-$R^2$ (p$R^2$). n.s.: not significant., *: $p < 0.05$, **: $p < 0.0001$ (Quantile test).

| Exp. | Description | Median K ($Y^{-1}$) | lower K ($Y^{-1}$) | upper K ($Y^{-1}$) | conf. int. width K ($Y^{-1}$) | p$R^2$ (CWI$_{PBOOT}$/CWI$_{FBOOT}$) |
|---|---|---|---|---|---|---|
| 1.1 | alba, ELEFAN_GA, PBoot | 2.03 | 1.63 | 2.97 | 1.34 | |
| 1.2 | alba, ELEFAN_GA, FBoot | 2.55 | 1.49 | 4.10 | 2.60 | 0.52** |
| 2.1 | alba, ELEFAN_SA, PBoot | 1.78 | 1.40 | 4.62 | 3.22 | |
| 2.2 | alba, ELEFAN_SA, FBoot | 2.68 | 1.07 | 4.88 | 3.81 | 0.85** |
| 3.1 | synLFQ5b, ELEFAN_GA, PBoot | 0.54 | 0.38 | 0.70 | 0.32 | |
| 3.2 | synLFQ5b, ELEFAN_GA, FBoot | 0.55 | 0.36 | 0.71 | 0.35 | 0.91* |
| 4.1 | synLFQ5b, ELEFAN_SA, PBoot | 0.49 | 0.27 | 0.60 | 0.32 | |
| 4.2 | synLFQ5b, ELEFAN_SA, FBoot | 0.49 | 0.27 | 0.60 | 0.32 | 1.0(n.s.) |

**Table 5: Results for the growth performance index (Phi') from eight bootstrap experiments (Exp 1.1. to Exp. 4.2) using the *Abra alba* and synlfq5b data sets.** N = 1,000 runs, search settings were optimized for precision (e.g., maxiter = 50, run = 10, popsize = 100, pmutation = 0.2 for ELEFAN_GA and SA_temp = 5e+05, maxit = 500 for ELEFAN_SA). Median, lower and upper 95% CIs, 95% confidence interval widths are given. The last column shows the pseudo-$R^2$ ($pR^2$). n.s.: not significant., *: $p < 0.05$, **: $p < 0.0001$ (Quantile test).

| Exp. | Description | Median Phi' | lower Phi' | upper Phi' | conf. int. width | $pR^2$ ($CIW_{PBoot}/CIW_{FBoot}$) |
|---|---|---|---|---|---|---|
| 1.1 | alba, ELEFAN_GA, PBoot | 2.31 | 2.25 | 2.46 | 0.21 | |
| 1.2 | alba, ELEFAN_GA, FBoot | 2.41 | 2.22 | 2.61 | 0.39 | 0.54** |
| 2.1 | alba, ELEFAN_SA, PBoot | 2.27 | 2.18 | 2.59 | 0.41 | |
| 2.2 | alba, ELEFAN_SA, FBoot | 2.41 | 2.09 | 2.69 | 0.60 | 0.68** |
| 3.1 | synLFQ5b, ELEFAN_GA, PBoot | 3.52 | 3.48 | 3.55 | 0.07 | |
| 3.2 | synLFQ5b, ELEFAN_GA, FBoot | 3.51 | 3.47 | 3.55 | 0.08 | 0.88** |
| 4.1 | synLFQ5b, ELEFAN_SA, PBoot | 3.50 | 3.46 | 3.53 | 0.07 | |
| 4.2 | synLFQ5b, ELEFAN_SA, FBoot | 3.50 | 3.46 | 3.53 | 0.07 | 1.0 (n.s.) |